# Magnetization switching of FePt nanoparticle recording medium by femtosecond laser pulses

R. John[1], M. Berritta[2], D. Hinzke[3], C. Müller[4], T. Santos[5], H. Ulrichs[6], P. Nieves[7,8], J. Walowski[1], R. Mondal[2], O. Chubykalo-Fesenko[7], J. McCord[4], P. M. Oppeneer[2], U. Nowak[3], M. Münzenberg[1,*]

[1] Department of Physics, Ernst-Moritz-Arndt-University, 17489 Greifswald, Greifswald, Germany.
[2] Department of Physics and Astronomy, Uppsala University, P.O. Box 516, SE-75120, Uppsala, Sweden.
[3] Department of Physics, University of Konstanz, 78457 Konstanz, Germany.
[4] Institute for Materials Science, Kiel University, 24143 Kiel, Germany.
[5] Western Digital Corporation, San Jose, CA 95138, USA.
[6] I. Phys. Institut, Georg-August-University, 37077 Göttingen, Germany.
[7] Instituto de Ciencia de Materiales de Madrid, CSIC, Cantoblanco, 28049 Madrid, Spain.
[8] International Research Center in Critical Raw Materials for Advanced Industrial Technologies, ICCRAM, Universidad de Burgos, 09001 Burgos, Spain.

**Manipulation of magnetization with ultrashort laser pulses is promising for information storage device applications. The dynamic of the magnetization response depends on the energy transfer from the photons to the spins during the initial laser excitation[1,2,3,4,5]. A material of special interest for magnetic storage is FePt nanoparticles[6], on which optical writing with optical angular momentum was demonstrated recently by Lambert *et al.*[7], although the mechanism remained unclear. Here we investigate experimentally and theoretically the all-optical switching of FePt nanoparticles. We show that the magnetization switching is a stochastic process. We develop a complete multiscale model which allows us to optimize the number of laser shots needed to write the magnetization of high anisotropy FePt nanoparticles in our experiments. We conclude that only angular momentum induced optically by the inverse Faraday effect will provide switching with one single femtosecond laser pulse.**

Since the first discovery of an ultrafast response of a spin system to a femtosecond laser pulse by Beaurepaire and colleagues[8], our understanding of how to use ultrashort laser pulses to control magnetization has increased considerably[9]. All-optical switching caused solely by the effect of an ultrashort laser pulse was demonstrated first for ferrimagnets[1,2,3,4], later for layered, synthetic ferrimagnets[5] and recently even for simple ferromagnets[7]. Importantly, two different kinds of all-optical (AOS) switching have to be distinguished, namely helicity-dependent all-optical switching (HD-AOS)[1,2,5], where the new magnetic orientation is defined by the optical angular momentum (helicity, of the circularly polarized laser light), and thermally driven switching caused by laser heating with linearly polarized light[3,4,10,11,12,13]. The latter has been observed in ferrimagnets only where the phenomenon has been connected with a transient ferromagnetic-like state, i.e., parallel alignment of the rare-earth and transition-metal sublattice magnetizations below the picosecond timescale[3,4,11,12]. Spin dynamics simulations[3,10] showed that this state follows from exchange of angular momentum between the antiparallel oriented moments on the two sublattices on a picosecond timescale. However, this mechanism does not apply to the HD-AOS observed for single lattice ferromagnets and consequently, the mechanisms underlying HD-AOS are currently under intensive debate[5,14,15,16,17]. It is evident that there must exist an asymmetry related to the helicity of the laser excitation which determines the probability of a switching event. The asymmetry in HD-AOS could originate from different absorptions of left and right circularly polarized light[18], i.e., a helicity-dependent thermal mechanism. Alternatively it could originate from the laser-induced magnetization caused by the helicity-dependent inverse Faraday effect (IFE)[9], essentially a non-thermal process. Both mechanisms rely on the very same optical transitions, and both originate from the interplay of spin-orbit coupling, exchange splitting and the helicity of the exciting laser field driving the transitions. Therefore, unveiling the microscopic origin of HD-AOS has been precluded so far. Here we combine measurements and multiscale simulations to come to the bottom of the HD-AOS in FePt.

We investigate FePt granular media designed for heat-assisted magnetic recording (HAMR)[6] with $\mu_0 H_S = 6\, T$ saturation field and employ magneto-optical Kerr effect (MOKE) microscopy on the macroscale of a few micrometres to record the magnetization switching. Fig. 1a shows the effect of writing using HD-AOS on FePt nanograins: starting with a randomly magnetized film, which means that 50% of the FePt grains are magnetized in 'up' and 50% magnetized in 'down' direction, with an average magnetization of zero, we find no magneto-optical contrast in Kerr effect images for writing with linear polarization, whereas for right (σ+) and left (σ-) circularly polarized pulses, we find a clear bright and dark contrast of the polar MOKE, respectively. This can be quantitatively analysed via cross sectional contrast profiles. We find a symmetric reversal starting with a 50%/ 50% ratio of up/ down magnetized grains (Fig. 1c). Starting with a 100%/ 0% ratio of up/ down magnetized grains we obtain writing probabilities of 63% and 41% for σ+ and σ- (Fig. 1d). Moreover, it is possible to write and overwrite the information starting with a 50%/ 50% ratio of up/ down magnetized FePt nanograins, as shown by two successive writing lines using first right (σ+) and then left (σ-) circularly polarized light in Fig. 1b. This demonstrates reversibility and hints at helicity as a source of the asymmetry. In addition, the observations point to a non-100% reversal for an infinite number of pulses that has to be understood.

Only multiscale calculations can combine information on the electronic level from *ab initio* calculations with the simulation of magnetization dynamics ranging from single FePt nanograins up to thermal macroscopic ensembles of thousands of particles. We start with *ab initio* calculations of the optical constants $n \pm$ for circularly polarized light and of the transient magnetization induced by the IFE. The former lead to helicity-dependent absorptions caused by the magnetic circular dichroism (MCD) that induce ultrafast heating. Taking both, the thermal effect and the imparted transient magnetization into account, a Landau-Lifshitz-Bloch- (LLB) type approach for a thermal spin ensemble allows us to calculate the switching probabilities of the FePt nanograins for a single laser pulse. Subsequently, we develop a rate model in which we employ these probabilities to derive analytic solutions for the magnetisation dynamics triggered by sequential shots. With that we discuss the conditions needed to realize 100%-one-shot switching. This provides a multiscale picture of the stochastic switching process that we compare to our measurements with sequential switching using repeated single laser pulses on FePt recording medium.

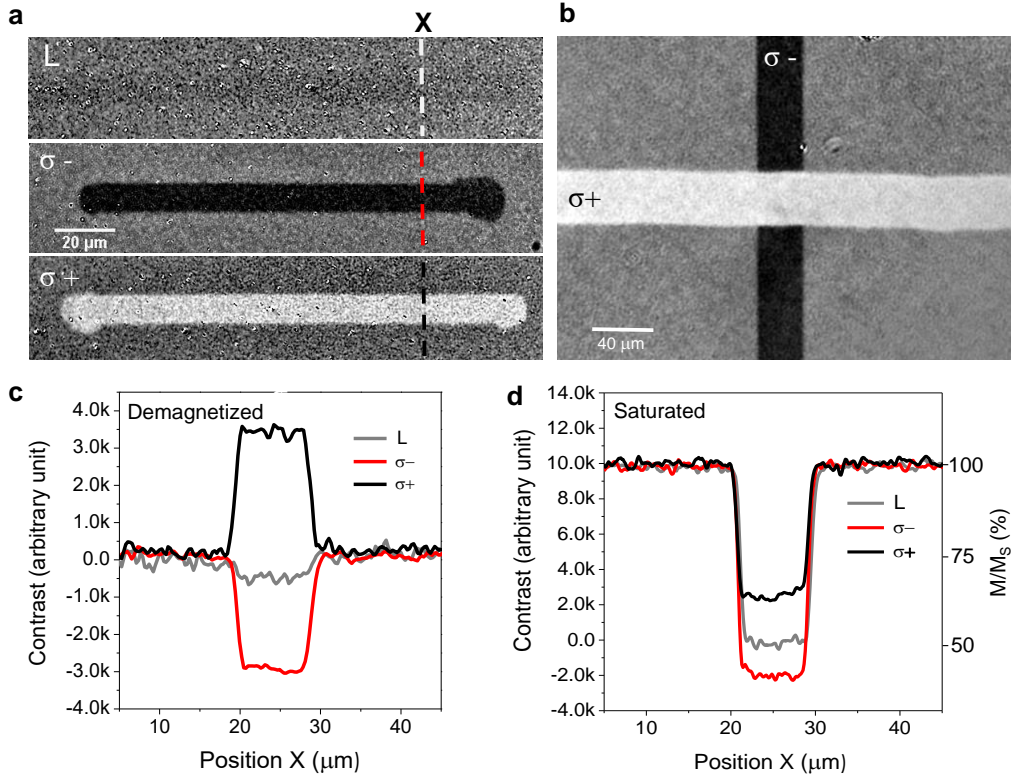

**Fig. 1. All-optical writing of a FePt recording medium. a,** Magneto-optical contrast images, starting with a demagnetized state one obtains a reversed magneto-optical contrast for opposite helicities (σ+, σ-) but not for linearly polarized light (L) along the line the laser spot has been moved. The number of laser pulses was here about 250 000 per spot. **b,** Overwriting of the magnetization direction is possible and reverses the magneto-optical contrast independently of the starting configuration. **c,** cross sectional contrast profiles along the dotted lines in **a,** starting with a demagnetized medium. **d,** Starting with a saturated medium with 100% $M_S$ gives a writing contrast of about 63% $M_S$ (σ+), 41% $M_S$ (σ-) (magneto-optical contrast image not shown). The average laser power onto the sample was 7.5 and 15 mW (6.6 and 13.2 mJ/cm$^2$ per pulse), respectively.

So far, models have been based on the existence of the IFE seen as a Raman-like optical transient state in dielectrics[19,20] or an internal field generated by the light field[2,21]. The strength of the effect, however, was never known and treated as a parameter. Differently from previous work, we calculate here directly and *ab initio* the magnetization that is induced in FePt through the optical angular momentum, driving the optical transitions, from recently derived expressions[22]. The IFE is a nonlinear optical effect related to electronic Raman and Rayleigh scattering processes. The central quantity is the induced helicity-dependent magnetization, which is given by

$$\Delta M_{ind}^{\sigma\pm}(\omega) = K_{IFE}^{\sigma\pm}(\omega) I/c \qquad (1)$$

where $K_{IFE}^{\sigma\pm}$ is the material, helicity and frequency-dependent IFE constant, *c* is the velocity of light and *I* is the laser intensity. The calculated IFE constants are given in Fig. 2a. In addition to a strong wave-length dependence that increases the induced magnetization for reduced photon energy, we also observe that, surprisingly, at the 1.55-eV photon energy used in the experiments, the helicity dependent induced magnetizations do not have opposite sign, as it would be if we had started with a paramagnetic material. Instead, in a ferromagnetic material the induced magnetization can have the same sign, but with a different amplitude. To calculate the amount of total magnetization induced, we multiply with the laser intensity. In our experiments, typical intensities range from 30 to 100 GW/cm², with peak intensities of up to 200 GW/cm² before absorption (see methods). The *ab initio* calculated values of $K_{IFE}^{\sigma-} = -0.033\,T^{-1}$ and of $K_{IFE}^{\sigma+} = -0.016\,T^{-1}$ at $\hbar\omega = 1.55\,eV$ and an light field intensity of 68 GW/cm² result in an induced magnetization of $\Delta M_{ind}^{\sigma-} = -0.22\,\mu_B$ and $\Delta M_{ind}^{\sigma+} = -0.10\,\mu_B$ per unit cell of FePt. Compared with the saturation magnetization, the size of laser-induced magnetization is small: it is about -7.1% and -3.5% of the saturation magnetization $M_S$, respectively. We further compute the helicity-dependent optical constants, $n\pm$, using $(n\pm)^2 = \varepsilon_{xx} \pm i\varepsilon_{xy}$, where $\varepsilon_{ij}(\omega)$ are elements of the *ab initio* calculated dielectric tensor. The imaginary part of $n\pm$ that determines the helicity-dependent absorption is shown in Fig. 2b. Due to the different absorptions caused by the MCD the increase of the electron temperature is asymmetric by about 40K at the peak electron temperature. As a remark, the IFE stems from the same optical transitions as the MCD and there is an absorptive contribution to the IFE[23]. But in contrast to the IFE, the MCD cannot induce any magnetization.

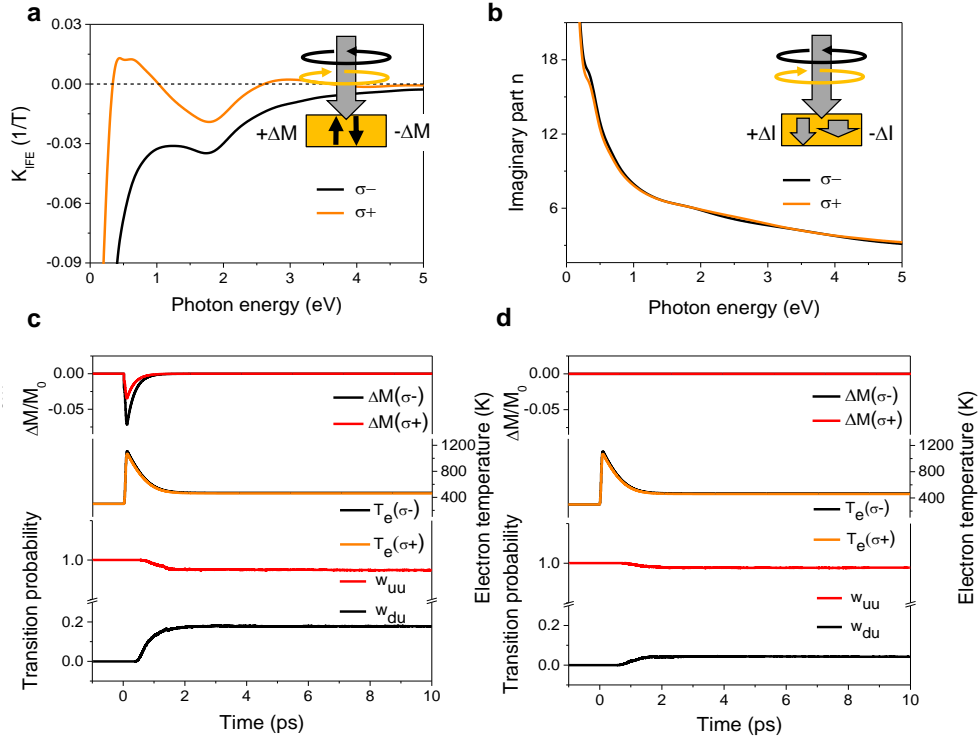

**Fig. 2. *Ab initio* calculations and switching probabilities. a**, The calculated inverse Faraday effect constant $K_{IFE}^{\sigma\pm}(\omega)$ of FePt for different photon energies $\hbar\omega$ and helicities σ±. **b**, Calculated imaginary part of the optical constant $n$ for different photon energies and helicities σ±. **c, d**, Magnetization switching in FePt, following a laser pulse triggering a sudden electron temperature rise with a peak electron temperature of about 1100K but with a slight difference due to the MCD (i.e., $T_e(\sigma\pm)$) at 1.55 eV of about 32 K, a peak inverse Faraday effect with a decay time of the IFE induced magnetization ΔM of -7.1% and -3.5% of the saturation magnetization $M_S$ of 250 fs. These parameters serve as an input for our magnetisation dynamics calculations using the LLB equation of motion. These calculations result in switching probabilities from 'down' to 'up', $w_{du}$, and 'up' to 'up', $w_{uu}$, in **c**, taking into account both IFE and MCD contributions, and in **d**, with the MCD only without IFE. The scenario corresponds to an average power onto the sample of 11mW (9.6 mJ/cm² per pulse).

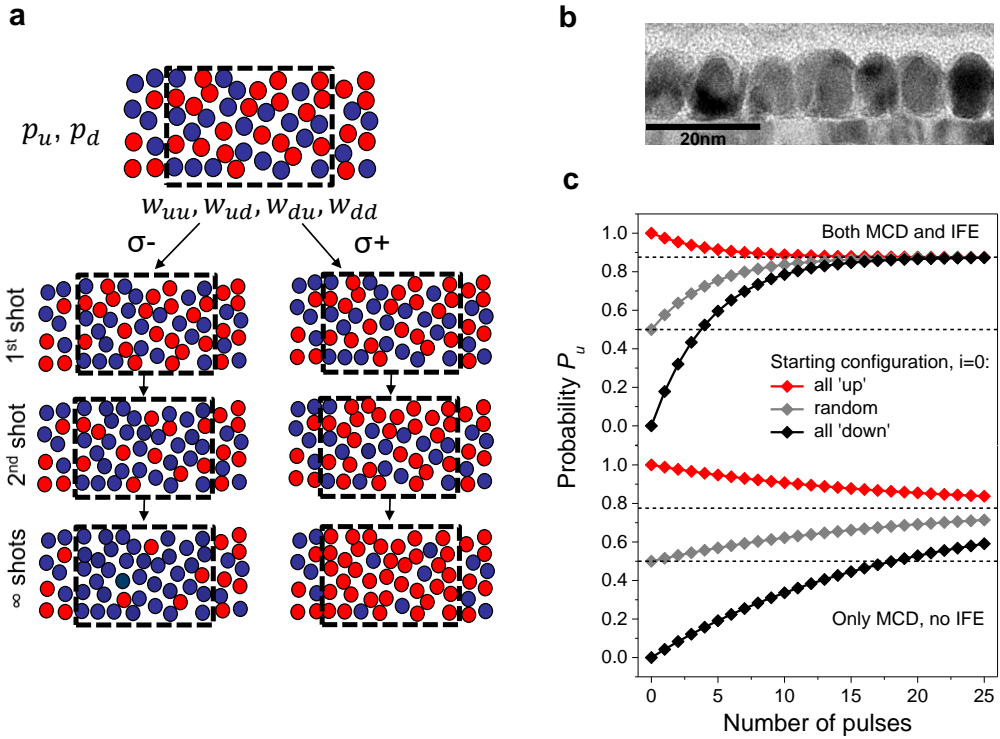

**Fig. 3. Microscopic structure and rate model. a,** Different switching probabilities lead to a final magnetisation of the FePt grain ensembles. The transition rates $w_{uu}, w_{ud}, w_{du}, w_{dd}$ determine the number of grains in the 'up' or 'down' states, described by the probabilities $p_u$ and $p_d$, after each single shot. **b,** Structure of the nanosize FePt grains: transmission electron micrograph showing the FePt grains on the seed layer. The grains have a coercive field of a few Tesla at room temperature, keeping them robust to thermal fluctuations. **c**, top: probability of being in an 'up' state ($P_u$) versus number of laser pulses starting from three different initial states when the helicity-dependent thermal heating via the MCD and the non-thermal influence of the IFE are taken into account. **c**, bottom: probability of being in an 'up' state versus number of laser pulses starting from three different initial states when only the MCD is taken into account, not the IFE. The scenario corresponds to an average power onto the sample of 11mW (9.6 mJ/cm$^2$ per pulse).

Our magnetization dynamics calculations are based on the stochastic LLB[24,25] equation with a single macro-spin per grain. The thermal input functions were calculated earlier within a multi-scale framework using an atomistic spin model that was based on an *ab initio* parameterization for FePt[26]. Specifically for FePt, the reduced electronic density of states near the Fermi energy causes heating of the electron system well above 1000 K, far above the Curie temperature, as shown earlier[27]. As a consequence, the FePt magnetisation approaches criticality and the grains might lose their magnetization information. This temperature rise, however, is slightly asymmetric because of a difference in the absorption of about ±2.5% for the two helicities. In addition to the sudden electron temperature rise, a small magnetization is induced by the IFE, present as an asymmetric magnetization contribution with a decay time which we assume slightly longer than the laser pulse itself (250 fs). All these quantities, which enter the magnetisation dynamics simulations are shown in the upper part of Fig. 2. Below, the resulting LLB dynamics is shown expressed as transition probabilities either to remain in the initial state (up–up) or to switch (down–up). The excitation pulse with right circular polarization always favors the up state (positive IFE). We calculate the dynamics for two scenarios, in Fig. 2c with IFE and MCD taken into account and, for comparison, in Fig. 2d with the MCD only. As a result we obtain different transition probabilities, for both cases, two of which are sufficient for the following rate theory, named $w_{uu}, w_{du}$, where $w_{uu}$ defines the probability for a transition from 'up' to 'up' and $w_{du}$ from 'down' to 'up'. These are employed in the rate model illustrated in Fig. 3a: because of the large anisotropy, one can assume in a good approximation a granular medium of decoupled, bistable FePt grains. They are either in 'up' or 'down' states with probabilities $p_u$ and $p_d$ in the ensemble. The magnetization is given by $M = M_S(T)(p_u - p_d)$. The thermal stochastic response is captured by four different transition probabilities, $w_{uu}, w_{ud}, w_{du}, w_{dd}$. They are related by $w_{uu} + w_{ud} = 1$ and $w_{du} + w_{dd} = 1$, so that only two transition probabilities are independent. The transition probabilities are determined via time integration of the LLB equation by taking into account the effects of heating, the IFE and the MCD. The nanoparticles cool sufficiently down between the pulses, so that we have blocked particles between events. Because of the total probability being $p_u + p_d = 1$, it is sufficient to discuss $\mathrm{P}u$ only. After one laser pulse the equation for the new probability is:

$$p_u^{i+1} = p_u^i w_{uu} + p_d^i w_{du} = w_{du} + p_d^i(w_{uu} - w_{du}) \tag{2}$$

We assume that the next event has identical transition probabilities. One can reformulate the combined probabilities as a geometrical series, and assuming *n* independent laser pulses one finds:

$$p_u^n = w_{du}\frac{(w_{uu} - w_{du})^n - 1}{w_{uu} - w_{du} - 1} + p^0(w_{uu} - w_{du})^n \tag{3}$$

Hence, the magnetization dynamics after successive laser pulses can be expressed in terms of the initial magnetization $\mathrm{P}^0$ and two transition probabilities, which are shown in the lower part of Fig. 2. The final state does not depend on the initial state but is simply given by the transition probabilities

$$p_u(n \to \infty) = w_{du}\frac{-1}{w_{uu} - w_{du} - 1} = \frac{w_{du}}{w_{ud} + w_{du}}. \tag{4}$$

We now discuss the consequences of the equation derived. Without any switching asymmetries, and for very high peak electron temperatures, FePt demagnetizes, which means that all transition rates become equal, $w_{uu} = 0.5, w_{ud} = 0.5$, and $p_u$ = 0.5, the demagnetized state. A low peak electron temperature, on the other hand, implies that no switching events occur, thus $w_{uu} = 1, w_{ud} = 0$. If we now implement the switching asymmetries, the IFE causes that, depending on helicity, either 'up' or 'down' is favoured. Assuming that 'up' is favoured we find $w_{uu} > w_{ud}$, but also $w_{du} > w_{ud} = 1 - w_{uu}$. Similarly, MCD leads to different degrees of heating of up- and down-magnetized FePt nanograins, so that the probabilities for switching are also asymmetric. This means that, in our rate model, the influences of IFE and MCD are not qualitatively distinguishable. However, these effects are still different, since only the IFE can reverse a magnetization. Thus quantitatively there will be differences in their efficiency: the perfect writing in the case of MCD would be a heating above Curie (or blocking) temperature of the down grains ($w_{du} = 0.5$), resulting in a random orientation, and no effect on the up grains ($w_{uu} = 1$), which would need about 5 to 10 pulses for writing. Conversely, for the perfect writing in the case of the IFE, we would need $w_{du} = 1$ and $w_{uu} = 1$, which is perfect writing in a single step. We thus predict from these two limiting cases that one-shot writing with a transition probability of 100% is only possible in the second case.

When the *ab initio* values are plugged in the Langevin spin dynamics simulation, the LLB-computed transition probabilities for the FePt nanograins we obtain are $w_{uu} = 0.86$ and $w_{du} = 0.38$. Plugging these numbers into the rate theory, we find that writing and rewriting with consecutive pulses are indeed possible. The resulting probabilities for multiple pulses are presented in Fig. 3c. After about 10 laser pulses $p_u$ converges to about 0.67, regardless of whether one starts with a fully polarized system ('up' or 'down') or a demagnetized system. This is in accord with our experimental findings.

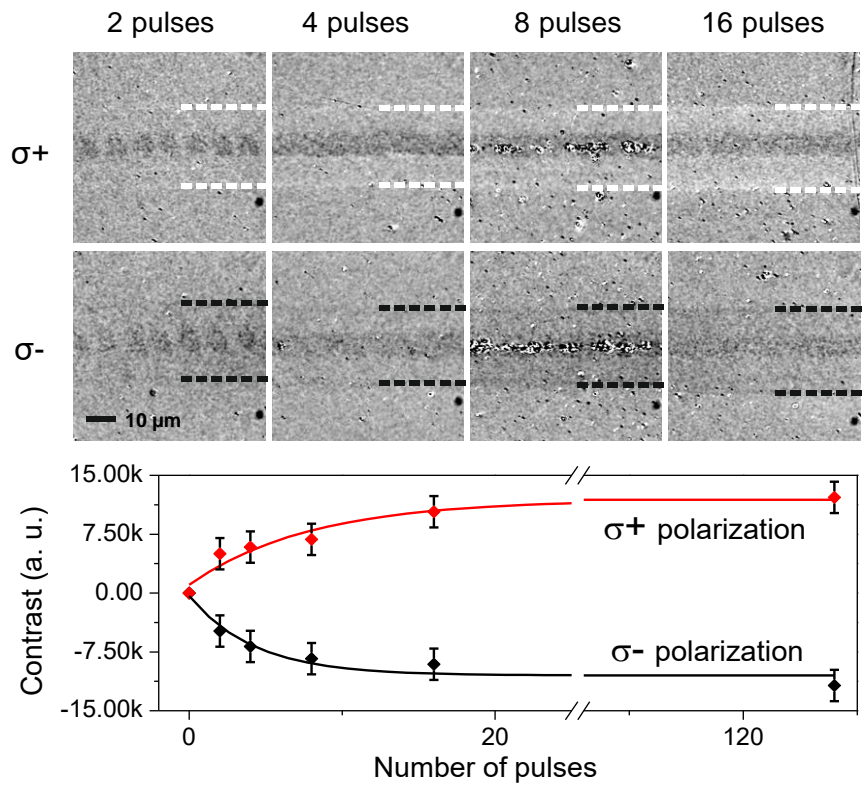

**Fig. 4. Magnetization switching experiments with consecutive single laser shots** starting from demagnetized recording media. Saturation is reached between 15 to 120 pulses of writing. Dark centre shows some excess heating and a structural modification of the FePt nanoparticles, which allows us to identify the pulse train distance. The number of pulses is indicated, the lines given in the bottom panel are exponential functions with a decay of 4.4(3) pulses. The average laser power was 5 mW onto the sample (30 mJ/cm² per pulse).

To compare the predictions of our rate theory, HD-AOS switching experiments using a varying number of subsequent pulses for writing were performed. Our results are shown in Fig. 4: the top row shows magneto-optical images using polar MOKE after the switching with σ+ helicity whereas the row below shows those obtained with σ- for a varying number of pulses. The average number of pulses per area was varied from 1 to 128, but only the images for up to 16 pulses are presented in Fig. 4. The central darker contrast is due to the modification and damage of the nanoparticles' carbon coating, in the centre of highest laser fluence. Yet this helps us to follow the pulse train, i.e. to visualize the average number of pulses over an area.

With an increasing number of pulses (from 2 to 16), i.e., from left to right panels, the magneto-optical contrast changes bright or dark for σ+ and σ-, respectively, with the accumulation of laser shots. In the area where switching is observed (Fig. 4), the fluence compared to the center fluence, is decreased by one half to below 15 mJ/cm$^2$. This fluence margin is well in accordance with our calculations. To analyse this quantitatively, similarly to the data in Fig.°1, we have taken the change of contrast as profiles along a line perpendicular to each piece of writing (varying the number of pulses) for both helicities and plotted the obtained contrast in the bottom panel. Our results support the claims of our rate theory for helicity-dependent AOS. Both the curves, experimental and theoretical calculation, show an accumulation of magnetization with each pulse increasing to a saturation rate.

From our combined experimental and theoretical investigation, we can unravel thermal and non-thermal contributions to the HD-AOS of FePt nanoparticles. We find that a principal difference between MCD and IFE assisted switching is that helicity-dependent heating via the MCD always leads only to a demagnetization stochastic processes and therefore cannot switch the magnetization deterministically. As a consequence, single shot will never be achieved with MCD. In contrast the IFE provides an additional magnetization contribution $\Delta M$ which could lead to a magnetization reversal if at the same time the thermal demagnetization leads to a nearly vanishing magnetization. Thus, as a consequence only the IFE can reverse the nanoparticle's magnetization with a single laser shot. A full multiscale approach leading to HD-AOS is required for a quantitative determination of the asymmetry parameters. Our approach allows the prediction of parameters for 100% switching with one shot for all-optical plasmonic write heads with polarization control that may address a single nanometer FePt grain, with a few 100 magnetic atoms, in future devices. Our work furthermore predicts how an optimization of the all-optical control of magnetism of FePt nanograins on femtosecond timescales can be achieved, with the central finding that optimized switching will be only possible by exploiting angular momentum induced via the IFE phenomenon.

**Materials and Methods**

**Fabrication**

FePt nanoparticles with $L1_0$ order and c-axis out-of-plane orientation were made by sputter deposition at elevated temperature[6]. The FePt grains are isolated by a non-magnetic segregant material at the grain boundaries and have a carbon overcoat protection layer on top. Hysteresis curves for the granular recording media reveal $\mu_0 H_S$ ~ 6 T and coercive field $\mu_0 H_C$ ~ 4 T.

**All-Optical Switching Using Ultrafast Laser Pulses**

We have performed AOS using the output of a Ti:Sapphire Regenerative Amplifier REGA 9040 (Coherent, Santa Clara, CA 95054, USA). The REGA was seeded by a Vitara Ti:Sapphire mode-locked oscillator which works at a frequency of 80 MHz. The pulse width (FWHM) after compressor REGA 9040 is measured to be 46 fs with a central wavelength of 800 nm. We determined about 60fs at the sample. The repetition rate of the laser after the amplifier was 250 kHz for writing/switching with a large number of pulses but was tuned down to 20 kHz for switching with a single/few pulses with the help of a chopper. The laser beam focused down to a beam waist of 17 µm in the first case and 23 µm in the second case. The average number of pulses over the switching area was varied by moving the sample at different speeds using a translation stage from Physik Instrumente GmbH.

**Magneto-optical Kerr effect microscopy**

Magneto-optical Kerr microscopy[28] with polar sensitivity has been realized in an adapted polarized light microscope (Zeiss Axio Imager) that is adapted for magnetic domain observations. Imaging was performed with a 50x objective with a numerical aperture NA = 0.8 and an illumination wavelength of $\lambda$ = 460 nm, resulting in a spatial resolution of approximately 300 nm. The weak magneto-optical contrast was enhanced by background subtraction of images with reversed magneto-optical contrast by switching between two different analyser angle settings in the microscope. Effects of spatially inhomogeneous illumination were compensated through a 2nd order polynomial surface intensity correction.

**Thermal modelling and internal light field**

A two-temperature model was used to determine the electron temperature induced by absorption of the light pulse in the opaque FePt. As before, we chose a specific set of material parameters for FePt, which assured consistency with the demagnetization dynamics observed in the time resolved MOKE and LLB modeling as described in [27]. In particular, the model was improved by using a Sommerfeld coefficient of $\gamma_e$ = 296.7 J/m$^3$K$^2$ derived *ab initio* from the density of states of FePt. A lattice heat capacity of $C_{ph}$ =1.0·10$^6$ J/m$^3$K, and an electron-phonon coupling constant of $G_{e-ph}$=4.0·10$^{17}$ W/m$^3$K had to be used to describe the temperature profiles. Our modelling shows that about 1.6% of the optical energy incident from outside is converted into heat in the FePt layer. In contrast, an optical transfer matrix calculation predicts a reflection of 70% of the light incident on the carbon protective layer, and a subsequent absorption of the remaining light in the FePt. This apparent contradiction can be explained by the granular structure of the FePt: assuming individual spherical particles, a rough estimate based on a Rayleigh-like absorption cross-section yields 0.8% absorption, which is close to the 1.6% found. For the calculation of the induced magnetization value by the IFE, the internal light field present in the FePt grains was used. The average power onto the film of 1mW equals 6.17·10$^9$ W/cm$^2$ local power density inside the FePt nanograins, using 21% of the total power and temporal shape of the 60fs laser pulse, a diameter of 17 µm and the repetition rate of 250 kHz (that includes 30% transmitted light through the carbon layer and the pulse shape).

**Magnetization dynamics calculations**

Our simulations are based on the stochastic LLB equation of motion[24,25] with a single macro-spin per grain. The necessary temperature dependent equilibrium properties (saturation magnetisation, exchange stiffness, parallel and perpendicular susceptibilities) were calculated earlier within a multi-scale framework[29] based on an atomistic spin model for FePt that was parameterized via *ab initio* methods[26]. As grain volume we assume (5nm)$^3$ and we simulate ensembles of 4096 non-interacting grains. The LLB dynamics describes the magnetic reaction to the thermal excitation (the electron temperature rise) and the IFE is considered as an additional contribution to the magnetization component perpendicular to the film with a decay time of 250fs. A saturation magnetization of 1050 kA/m was used. At any time during the simulation, transition probabilities can be calculated as relative number of grains where the perpendicular component of the magnetization has switched sign.

## Acknowledgements

This work was supported by the EC under Contract No. 281043, FemtoSpin. The work at Greifswald University was supported by the German research foundation (DFG), projects MU MU 1780/ 8-1, MU 1780/ 10-1. Research at Göttingen University was supported via SFB 1073, Projects A2 and B1. Research at Uppsala University was supported by the Swedish Research Council (VR), the Röntgen-Ångström Cluster, the Knut and Alice Wallenberg Foundation (Contract No. 2015.0060), and Swedish National Infrastructure for Computing (SNIC). Research at Kiel University was supported by the DFG, projects MC 9/ 9-2, MC 9/ 10-2. P.N. acknowledges support from EU Horizon 2020 Framework Programme for Research and Innovation (2014-2020) under Grant Agreement No. 686056, NOVAMAG. The work in Konstanz was supported via the Center for Applied Photonics.

## Author contributions

Experimental idea was designed by M.M., T.S., O.C.-F., P.M.O., U.N., all-optical switching, imaging and image analysis was conducted by R.J., C.M., J.W., M.M., J.M., samples were prepared by T.S., thermal model parameters were analyzed by H.U., P.N., O.C.-F., M.M., development and calculation of IFE and MCD were conducted by M.B., R.M., P.M.O., modeling of LLB dynamics D.H., P.N., O.C.-F., U.N., and rate model was developed by D.H., U.N., manuscript was written by R.J., O.C.-F., P.M.O., U.N., M.M., all authors discussed the models and the manuscript.